\documentclass[aps,showpacs,floatfix]{revtex4}
\usepackage{amssymb}
\usepackage{amsmath}
\usepackage{epsfig}
\usepackage{subfigure}
\usepackage{multirow}

\begin{document}

\title{ The Existence of G\"{o}del, Einstein and de Sitter Universes}
\author{Timothy Clifton}
\email{T.Clifton@damtp.cam.ac.uk}
\affiliation{DAMTP, Centre for Mathematical Sciences, University of Cambridge,
Wilberforce Road, Cambridge, CB3 0WA, UK}
\author{John D. Barrow}
\email{J.D.Barrow@damtp.cam.ac.uk}
\affiliation{DAMTP, Centre for Mathematical Sciences, University of Cambridge,
Wilberforce Road, Cambridge, CB3 0WA, UK}
\date{\today }

\begin{abstract}
We determine the general conditions for the existence of G\"{o}del, Einstein
static, and de Sitter universes in gravity theories derived from a
Lagrangian that is an arbitrary function of the scalar curvature and Ricci
and Riemann curvature invariants. Explicit expressions for the solutions are
found in terms of the parameters defining the Lagrangian. We also determine
the conditions on the Lagrangian of the theory under which time-travel is
allowed in the G\"{o}del universes. 
\end{abstract}

\maketitle

\section{Introduction}

We consider the conditions required for gravity theories that are derived
from Lagrangians that are functions of the scalar curvature and Ricci and
Riemann curvature invariants to possess solutions which are homogeneous
space-times of the G\"{o}del, Einstein static, and de Sitter forms. In the G%
\"{o}del case we determine the conditions for the existence or non-existence
of closed time-like curves in these universes. These three homogeneous
space-times, and the investigations of their stability, have played a
central role in our understanding of the dynamics of general relativity and
in the possible astrophysical consequences of general relativistic
cosmologies. Higher-order modifications of general relativity are of
importance in assessing the corrections that might be introduced to general
relativity in high-curvature environments and can also be of use in
explaining the late-time acceleration of the universe. Furthermore,
investigations of these theories allows for an evaluation of the special
nature of general relativity itself. With this in mind we have recently
provided a detailed analysis of the cosmological consequences of gravity
theories with power-law Lagrangians \cite{Cli05, Bar05}. Here, we extend and
broaden this study to include the conditions under which the G\"{o}del,
Einstein static, and de Sitter universes exist in a wider class of
non-linear gravity theories.

In section \ref{fieldequations} we present the relevant field equations for
the gravitational theories that we will be considering. In section \ref%
{godeluniverses} we find the conditions for the existence of G\"{o}del
universes in these theories and perform an analysis of the conditions for
the existence of closed time-like curves in these solutions. Sections \ref%
{einsteinstatic} and \ref{desitter} contain the conditions for the existence
of Einstein static and de Sitter universes, respectively, and in section \ref%
{conclusions} we summarize our results.

\section{Field equations}

\label{fieldequations}

In this paper we consider gravitation theories derived from a function of
the three possible linear and quadratic contractions of the Riemann
curvature tensor; $R$, $R_{ab}R^{ab}$ and $R_{abcd}R^{abcd}$ \cite{footnote1}%
. The relevant weight-zero scalar density for this general class of theories
is then given by 
\begin{equation}
\mathcal{L}_{G}=\chi ^{-1}\sqrt{g}f(X,Y,Z)  \label{density}
\end{equation}%
where $f(X,Y,Z)$ is an arbitrary function of $X$, $Y$ and $Z$ which are
defined by $X=R$, $Y=R_{ab}R^{ab}$ and $Z=R_{abcd}R^{abcd}$; $\chi $ is an
arbitrary constant which can be determined by the appropriate Newtonian
low-curvature limit. The action is obtained, as usual, by integrating this
density together with that of the matter fields over all space. The addition
of supplementary terms to the density (\ref{density}) in order to cancel
total divergences which can be transformed to integrals on the boundary can
be problematic (see e.g. \cite{Mad88}) and so, for simplicity, they will all
be assumed to vanish.

Taking the first variation of the action derived from (\ref{density}),
together with that for the matter fields and a cosmological constant, then
gives 
\begin{equation*}
\delta I = \chi^{-1} \int d\Omega (-P^{a b}-g^{a b} \Lambda+ \frac{\chi}{%
2} T^{a b}) \delta g_{ab}
\end{equation*}
where 
\begin{multline}  \label{P}
P^{a b} = -\frac{1}{2} f g^{a b} + f_X R^{a b}+2 f_Y R^{c (a} {R^{b)}}_{c}+2
f_Z R^{e d c (a} {R^{b)}}_{c d e} +f_{X; c d}(g^{a b} g^{c d}-g^{a c} g^{b d})
\\
+\square (f_Y R^{a b}) + g^{a b} (f_Y R^{c d})_{;c d}-2 (f_Y R^{c (a})_{;\;
\; c}^{\; b)}-4 (f_Z R^{d (a b) c})_{;c d}.
\end{multline}
Here, $\Lambda$ is the cosmological constant (defined independent of $%
f(X,Y,Z)$) and $T^{a b}$ is the energy-momentum tensor of the matter. The
notation $f_N$ denotes partial differentiation of $f$ with respect to
$N$.  A derivation of equation (\ref{P}) is given in the appendix.
Looking for a stationary point of the action requires setting the first
variation to zero, giving the field equations 
\begin{equation}  \label{fequations}
P_{a b}=\frac{\chi}{2} T_{a b} - g_{a b} \Lambda.
\end{equation}
These field equations are generally of fourth-order, with the exception of
the cases in which the function $f$ is linear in the second derivatives of
the metric, which notably includes general relativity where $f=X$. This
property makes these equations particularly difficult to solve. Considerable
simplification occurs if we assume that the three curvature scalars $X$, $Y$
and $Z$ are constant. In this case, the expression (\ref{P}) reduces to 
\begin{equation}
P^{ab}=-\frac{1}{2}%
fg^{ab}+R^{ab}(f_{X}-2Rf_{Z})-2R^{acdb}R_{cd}(f_{Y}+4f_{Z})+\frac{1}{2}%
g^{ab}(X^{2}-4Y+Z)f_{Z}  \label{P2}
\end{equation}%
where use has been made of the identities \cite{DeW65} 
\begin{align*}
{R^{abcd}}_{;bc}& =-\square R^{ad}+{{{R^{ac}}_{;}}^{d}}_{c} \\
{{{R^{ac}}_{;}}^{d}}_{c}& =\frac{1}{2}R_{;}^{\;ad}+R^{abed}R_{be}+R_{%
\;c}^{a}R^{dc} \\
{R^{ab}}_{;ab}& =\frac{1}{2}\square R \\
2{R^{a}}_{cde}R^{bcde}+2g^{ab}Y+2RR^{ab}& =\frac{1}{2}%
g^{ab}Z-4R^{acdb}R_{cd}+4{R^{a}}_{c}R^{bc}+\frac{1}{2}g^{ab}R^{2}.
\end{align*}%
Equation (\ref{P2}) is only of second order and is therefore a significant
simplification of the original system of equations. It is the solutions of
these equations that we will now study.

There are a number of highly symmetric space-times in which the curvature
scalars $X$, $Y$ and $Z$ take constant values, including the G\"{o}del \cite%
{God49, heck, tsag}, Einstein static \cite{ellis}, and de Sitter universes 
\cite{HE}. We will investigate solutions of the field equations specified by
(\ref{P2}) for these three homogeneous space-times.

\section{G\"{o}del universes}

\label{godeluniverses}

The G\"{o}del universe \cite{God49, heck, tsag} is a homogeneous space-time
given by the line element 
\begin{align}
ds^{2}& =-(dt+C(r)d\psi )^{2}+D^{2}(r)d\psi ^{2}+dr^{2}+dz^{2}  \label{godel}
\\
& =-dt^{2}-2C(r)dtd\psi +G(r)d\psi ^{2}+dr^{2}+dz^{2} \nonumber
\end{align}
where 
\begin{align*}
C(r)& =\frac{4\Omega }{m^{2}}\sinh ^{2}\left( \frac{mr}{2}\right) \\
D(r)& =\frac{1}{m}\sinh (mr) \\
G(r)& =D^{2}(r)-C^{2}(r) \\
& =\frac{4}{m^{2}}\sinh ^{2}\left( \frac{mr}{2}\right) \left[ 1+\left( 1- 
\frac{4\Omega ^{2}}{m^{2}}\right) \sinh ^{2}\left( \frac{mr}{2}\right) %
\right]
\end{align*}
and $m$ and $\Omega $ are constants. The existence of G\"{o}del universes in
one particular $f(X,Y)$ theory has been previously studied by Accioly \cite%
{Acc87} where a solution was found in vacuum. We extend this analysis to the
more general class of theories above and to universes filled with matter
fluids.

The G\"{o}del universe is of particular theoretical interest as it allows
the possibility of closed time-like curves, and hence time-travel \cite{bt1,
bt2}. The condition required to avoid the existence of closed time-like
curves is \cite{reb, bd} 
\begin{equation*}
G(r)>0\qquad \text{for}\qquad r>0
\end{equation*}%
or 
\begin{equation}
m^{2}\geqslant 4\Omega ^{2}.  \label{ineq}
\end{equation}%
By investigating the existence of G\"{o}del universes in this general class
of gravity theories we will also be able to determine those theories in
which the time-travel condition (\ref{ineq}) is satisfied. Consequently, we
will be able to determine those theories in which time travel is a
theoretical possibility.

It is convenient to work in the non-holonomic basis defined by 
\begin{align*}
{e^{(0)}}_0 &= {e^{(2)}}_2 = {e^{(3)}}_3 = 1 \\
{e^{(1)}}_1 &= D(r) \\
{e^{(0)}}_1 &= C(r).
\end{align*}
The line-element (\ref{godel}) then becomes 
\begin{equation*}
ds^2=-(\theta^{(0)})^2+(\theta^{(1)})^2+(\theta^{(2)})^2+(\theta^{(3)})^2
\end{equation*}
where the one-forms $\theta^{(A)}$ are given by $\theta^{(A)}=e^{(A)}_{\quad
a}dx^a$ (capital Latin letters denote tetrad indices and lower case Latin
letters denote space-time indices). The inverses of $e^{(A)}_{\quad a}$ can
be calculated from the relations $e^{(A)}_{\quad a} e^{a}_{\; (B)}={%
\delta^{(A)}}_{(B)}$ and the non-zero elements of the Riemann tensor in this
basis are then given by 
\begin{align*}
R_{(0)(1)(0)(1)} &= R_{(0)(2)(0)(2)} = \Omega^2 \\
R_{(1)(2)(1)(2)} &= 3 \Omega^2 - m^2.
\end{align*}

The perfect-fluid energy-momentum tensor is defined in the usual way with
respect to the comoving 4-velocity $U^{a}=(1,0,0,0)$ and its covariant
counterpart $U_{a}=(-1,0,-C(r),0)$ such that its non-zero components in the
non-holonomic basis are given by 
\begin{equation*}
T_{(0)(0)}=\rho \qquad \text{and}\qquad T_{(1)(1)}=T_{(2)(2)}=T_{(3)(3)}=p.
\end{equation*}

The field equations, (\ref{fequations}), for this space-time can then be
manipulated into the form 
\begin{align}
\Lambda -\frac{\chi }{2}p& =\frac{1}{2}f  \label{1} \\
0& =(2\Omega ^{2}-m^{2})(f_{X}-4[\Omega
^{2}-m^{2}]f_{Z})+2(f_{Y}+4f_{Z})(8\Omega ^{4}-5\Omega ^{2}m^{2}+m^{4})
\label{1a} \\
\frac{\chi }{2}(\rho +p)& =2\Omega ^{2}(f_{X}-4[\Omega
^{2}-m^{2}]f_{Z})+4(f_{Y}+4f_{Z})\Omega ^{2}(2\Omega ^{2}-m^{2}).  \label{1b}
\end{align}%
Solving the field equations has now been reduced to solving these three
algebraic relations for some specified $f(X,Y,Z)$.

\subsection{$f=f(X)$}

For $\rho +p\neq 0$ we see from (\ref{1b}) that $f_{X}\neq 0$. From equation
(\ref{1a}) it can then be seen that $m^{2}=2\Omega ^{2}$, as in general
relativity. Therefore for any theory of the type $f=f(X)$ the inequality (%
\ref{ineq}) is not satisfied and closed time-like curves exist, when $\rho
+p\neq 0$.

It now remains to investigate the case $\rho +p=0$. It can immediately be
seen from (\ref{1b}) that we must have $f_{X}=0$ in order for a solution to
exist. This sets the relation between $m$ and $\Omega $ and automatically
satisfies equation (\ref{1a}). The required value of $\Lambda $ can then be
read off from equation (\ref{1}). The condition $f_{X}=0$ will now be
investigated for a variety of specific theories.

\subsubsection{$f=f(X)=X+\protect\alpha X^2$}

Theories of this kind have been much studied \cite{BO, ker} as they have a
number of interesting properties, not least of which is that they display
divergences which are normalisable at the one loop level \cite{Ste77}.

In these theories, the condition $f_{X}=0$ is equivalent to 
\begin{equation*}
m^{2}=\Omega ^{2}+\frac{1}{4\alpha }.
\end{equation*}%
The condition under which this theory then satisfies the inequality (\ref%
{ineq}), and hence does not admit closed time-like curves, is $\Omega^2
\leqslant (12 \alpha)^{-1}$. Therefore, for any given theory of this kind,
defined only by a choice of the constant $\alpha $, there is a range of
values of $\Omega $ for which closed time-like curves do not exist, when $%
\alpha>0$ and $\rho +p=0$. However, when $\alpha<0$ this condition is never
satisfied and closed time-like curves exist for all values of $\Omega$.

\subsubsection{$f=f(X)=X+\frac{\protect\alpha^2}{X}$}

Theories of this type have generated considerable interest as they introduce
cosmological effects at late times, when $R$ is small, which may be able to
mimic the effects of dark energy on the Hubble flow \cite{Car05,Noj03, Nav}.
The square in the factor $\alpha ^{2}$ is introduced here as these theories
require a positive value for this coefficient in order for the field
equations to have a solution.

When $f_{X}=0$ we have the two possible relations 
\begin{equation*}
m^{2}=\Omega ^{2}\pm \frac{\alpha }{2}
\end{equation*}%
where $\alpha $ is the positive real root of $\alpha ^{2}$. The upper branch
of this solution then allows the condition (\ref{ineq}) to be satisfied if $%
6\Omega ^{2}\leqslant \alpha $; or, for the lower branch, if $6\Omega
^{2}\leqslant -\alpha $. For any particular value of $\alpha >0$, the upper
branch always admits a range of $\Omega $ for which closed time-like curves
do not exist. For the lower branch, however, the inequality (\ref{ineq}) is
never satisfied and closed time-like curves are permitted for any value of $%
\Omega $.

\subsubsection{$f=f(X)=\vert X \vert^{1+\protect\delta}$}

This scale-invariant class of theories is of interest as its particularly
simple form allows a number of physically relevant exact solutions to be
found \cite{Cli05,Bar05,schm,duns}. In order for solutions of the G\"{o}del
type to exist in these theories we must impose upon $\delta $ the constraint 
$\delta \geqslant 0$.

The condition $f_{X}=0$ now gives 
\begin{equation*}
m^{2}=\Omega ^{2}.
\end{equation*}%
Evidently, the condition (\ref{ineq}) is always satisfied in this case: G%
\"{o}del solutions always exist and closed time-like curves are permitted
for any value of the vorticity parameter $\Omega $.

\subsection{$f=f(X,Y,Z)$}

In this general case, equations (\ref{1a}) and (\ref{1b}) can be manipulated
into the form 
\begin{align}
f_{X}-4(\Omega ^{2}-m^{2})f_{Z}& =\frac{(8\Omega ^{4}-5\Omega
^{2}m^{2}+m^{4})}{4\Omega ^{2}(4\Omega ^{2}-m^{2})}\chi (\rho +p)  \label{4}
\\
f_{Y}+4f_{Z}& =-\frac{(2\Omega ^{2}-m^{2})}{8\Omega ^{2}(4\Omega ^{2}-m^{2})}%
\chi (\rho +p)  \label{5}
\end{align}%
when $m^{2}\neq 4\Omega ^{2}$. The special case $m^{2}=4\Omega ^{2}$ gives 
\begin{align}
\rho +p& =0 \\
f_{X} &= 4 \Omega^2 (f_Y+f_Z).
\end{align}

First we consider $\rho +p>0$. It is clear that $m^{2}=4\Omega ^{2}$ is not
a solution in this case. Equations (\ref{4}) and (\ref{5}) show that in
order to satisfy the inequality (\ref{ineq}), and avoid the existence of
closed time-like curves, the following two conditions must be satisfied
simultaneously 
\begin{align}
f_{X}+4(m^{2}-\Omega ^{2})f_{Z}& <0  \label{G2} \\
f_{Y}+4f_{Z}& <0.  \label{G3}
\end{align}%
For $\rho +p<0$ the inequalities must be reversed in these two equations. It
is now clear that if it is possible to construct a theory which has a
solution of the G\"{o}del type without closed time-like curves for a
non-zero $\rho +p$ of a given sign, then this theory cannot ensure the
non-existence of closed time-like curves for the opposite sign of $\rho +p$.

It remains to investigate the case $\rho+p=0$. In order to have a solution
in this case we require that the two equations 
\begin{align}
f_X-4 (\Omega^2-m^2) f_Z &= 0 \\
f_Y+4 f_Z &= 0
\end{align}
are simultaneously satisfied, for $m^2 \neq 4 \Omega^2$, or 
\begin{equation}  \label{G4}
f_X = 4 \Omega^2 (f_Y+f_Z)
\end{equation}
for $m^2 = 4 \Omega^2$. Any theory which satisfies (\ref{G4}), therefore,
does not allow the existence of closed time-like curves when $\rho+p=0$.

We now illustrate these considerations by example.

\subsubsection{$f=X+ \protect\alpha X^2 +\protect\beta Y +\protect\gamma Z$}

Using the identity \cite{DeW65} 
\begin{equation*}
\frac{\delta }{\delta g_{ab}}\left( \sqrt{g}(X^{2}-4Y+Z)\right) =\text{pure
divergence}
\end{equation*}%
it is possible to rewrite $f$ as $f=X+\hat{\alpha}X^{2}+\hat{\beta}Y$ where $%
\hat{\alpha}=\alpha -\gamma $ and $\hat{\beta}=\beta +4\gamma $. This is the
theory considered by Accioly \cite{Acc87}.

For the case $\rho +p>0,$ the inequalities (\ref{G2}) and (\ref{G3}) then
become 
\begin{equation}
\hat{\beta}<0\qquad \text{and}\qquad \hat{\alpha}>\frac{1}{2|R|}.
\label{ineq2}
\end{equation}%
The first of these inequalities can be satisfied trivially. The second can
be only be satisfied for all $\Omega $ if $|R|$ has some non-zero minimum
value. From the field equations (\ref{1a}) and (\ref{1b}) we can obtain 
\begin{equation*}
(2\Omega ^{2}-m^{2})(1+4\hat{\alpha}[\Omega ^{2}-m^{2}])=-2\hat{\beta}%
(8\Omega ^{4}-5\Omega ^{2}m^{2}+m^{4})
\end{equation*}%
which gives two solutions for $m$ as a function of $\Omega $, $\hat{\alpha}$
and $\hat{\beta}$. Substituting either of these values of $m$ into $%
R=2(\Omega ^{2}-m^{2})$ then gives that $R\rightarrow 0$ as $\Omega
^{2}\rightarrow \frac{1}{8 \vert \hat{\beta} \vert}$. It can now be seen that the second
equality in (\ref{ineq2}) cannot be satisfied for all $\Omega $ if $\hat{%
\alpha}$ is finite. It is, therefore, not possible to construct a theory of
this type which excludes the possibility of closed time-like curves for all $%
\Omega $, when $\rho +p\neq 0$.

For $\rho+p<0$ the inequalities in (\ref{ineq2}) must be reversed. In this
case $\hat{\beta}>0$ and $\hat{\alpha}>0$ allows a range of $\Omega$ in
which closed time-like curves are not permitted and $\hat{\beta}>0$ and $%
\hat{\alpha}<0$ does not allow closed time-like curves for any values of $%
\Omega$.

The case $\rho +p=0$ was studied by Accioly \cite{Acc87} where it was found
that the equation (\ref{G4}) is satisfied if 
\begin{equation*}
4\Omega ^{2}=m^{2}=\frac{1}{(3\hat{\alpha}+\hat{\beta})},
\end{equation*}%
hence closed time-like curves do not exist in G\"{o}del universes for these
theories, when $\rho +p=0$.

\subsubsection{$f=\protect\alpha X+\protect\beta Y$}

This class of theories is introduced as an example which excludes the
possibility of closed time-like curves for $\rho+p>0$. The inequalities (\ref%
{G2}) and (\ref{G3}) in this case are 
\begin{equation*}
\alpha <0 \qquad \text{and} \qquad \beta <0
\end{equation*}
which can be trivially satisfied. For $\rho+p=0$ the only non-trivial
solution is given by $m^2 = 4 \Omega^2$, the value of $\Omega$ then being
given in terms of $\alpha$ and $\beta$ by (\ref{G4}).

This example shows explicitly that it is possible to construct a theory in
which closed time-like curves do not occur in G\"{o}del universes when $\rho
+p\geqslant 0$ (though we do not consider it as physically viable as $\alpha 
$ is required to have the `wrong' sign \cite{BO, Ruz}).

Table \ref{table} summarises the results found in this section.

\begin{table}[tbp]
\par
\begin{center}
\begin{tabular}{|c|c|c|c|c|}
\hline
\multirow{2}{*}{$f(X,Y,Z)$} & Additional & \multicolumn{3}{|c|}{Closed
time-like curves exist when} \\ \cline{3-5}
& conditions & $\; \rho+p > 0 \;$ & $\; \rho+p < 0 \;$ & $\; \rho+p = 0 \;$
\\ \hline
$X+ \alpha X^2$ & $\alpha >0$ & $\checkmark$ & $\checkmark$ & $\checkmark$/$%
\times$ \\ 
& $\alpha <0$ & $\checkmark$ & $\checkmark$ & $\checkmark$ \\ 
&  &  &  &  \\ 
$X+\frac{\alpha^2}{X} $ & $+$ve branch & $\checkmark$ & $\checkmark$ & $%
\checkmark$/$\times$ \\ 
& $-$ve branch & $\checkmark$ & $\checkmark$ & $\checkmark$ \\ 
&  &  &  &  \\ 
$\vert X \vert^{1+\delta}$ & $\delta >0$ & $\checkmark$ & $\checkmark$ & $%
\checkmark$ \\ 
&  &  &  &  \\ 
$\; X+\alpha X^2 + \beta Y + \gamma Z \;$ & $\; \alpha-\gamma >0$, $\; \beta
+ 4 \gamma<0 \;$ & $\checkmark$/$\times$ & $\checkmark$ & $\times$ \\ 
& $\alpha-\gamma <0$, $\; \beta + 4 \gamma<0$ & $\checkmark$ & $\checkmark$
& $\times$ \\ 
& $\alpha-\gamma >0$, $\; \beta + 4 \gamma>0$ & $\checkmark$ & $\checkmark$/$%
\times$ & $\times$ \\ 
& $\alpha-\gamma <0$, $\; \beta + 4 \gamma>0$ & $\checkmark$ & $\times$ & $%
\times$ \\ 
&  &  &  &  \\ 
$\alpha X + \beta Y$ & $\alpha >0$, $\; \beta <0$ & $\checkmark$ & $%
\checkmark$ & $\times$ \\ 
& $\alpha >0$, $\; \beta >0$ & $\checkmark$ & $\times$ & $\times$ \\ 
& $\alpha <0$, $\; \beta <0$ & $\times$ & $\checkmark$ & $\times$ \\ 
& $\alpha <0$, $\; \beta >0$ & $\checkmark$ & $\checkmark$ & $\times$ \\ 
\hline
\end{tabular}%
\end{center}
\caption{A summary of the conditions under which closed time-like curves can
exist in G\"{o}del universes, for various different gravitational theories,
defined by $f(X,Y,Z)$: $\checkmark $ denotes their existence for all values
of $\Omega $ and $\times $ denotes that they are not allowed for any value
of $\Omega $. The symbol $\checkmark $/$\times $ means that closed time-like
curves are allowed to exist for some restricted range of $\Omega $ only; the
ranges are given in the main text.}
\label{table}
\end{table}

\section{Einstein static universes}

\label{einsteinstatic}

The Einstein static universe is a homogeneous and isotropic space-time with
line element 
\begin{equation*}
ds^{2}=-dt^{2}+\frac{dr^{2}}{(1-\kappa r^{2})}+r^{2}(d\theta ^{2}+\sin
^{2}\theta d\phi ^{2}).
\end{equation*}%
Here, $\kappa $ parametrizes the curvature of the space-like slices
orthogonal to $t$ and the scale-factor has been rescaled to 1. For a
universe containing pressureless dust the field equations (\ref{fequations})
can now be written as 
\begin{align} 
\chi \rho & =4\kappa f_{X}+16\kappa ^{2} (f_{Y}+f_{Z})
\label{static1} \\
 \label{static2}
\Lambda & =\frac{1}{2}f-2\kappa f_{X}-8\kappa ^{2} (f_{Y}+f_{Z}).
\end{align}%
It can now be seen immediately that solutions exist for an Einstein static
universe for any $f(X,Y,Z)$ that is differentiable in all its arguments. The
corresponding values of $\rho $ and $\Lambda $ are simply read off from
equations (\ref{static1}) and (\ref{static2}). It remains to be studied
under what circumstances these solutions are stable. The investigation by
Barrow, Ellis, Maartens, and Tsagas \cite{ellis} shows that this is an issue
that depends upon the material content and the equation of state of matter
in a delicate fashion. In general relativity, there is first-order stability
against density perturbations when the sound speed exceeds a critical value (%
$1/\sqrt{5}$ of the speed of light) because the Jeans length exceeds the
size of the universe \cite{gibb, ellis}. However, there is instability
against homogeneous gravitational-wave modes of Bianchi IX type \cite{ellis}%
. In general, we expect that a universe with compact space sections and
Killing vectors to display linearisation instability \cite{BT}. The function
space of general solutions to Einstein's equations possesses a conical
structure at these particular special solutions and there are an infinite
number of perturbation expansions tangential to the conical point which do
not converge to true solutions to the field equations. This is only ensured
if further constraints are satisfied and some investigation of these
problems was made by Losic\ and Unruh \cite{unruh}. Similar studies could be
performed for the new family of solutions that we have identified here.

\section{De Sitter universes}

\label{desitter}

The line element for the maximally symmetric de Sitter vacuum universe can
be written as 
\begin{equation*}
ds^{2}=-dt^{2}+e^{2\sqrt{\frac{\Lambda }{3}}t}(dr^{2}+r^{2}(d\theta
^{2}+\sin ^{2}\theta d\phi ^{2}))
\end{equation*}%
where $\Lambda $ is the cosmological constant. For de Sitter space-time all
components of the Riemann and Ricci tensors can be written in terms of the
Ricci scalar using the equations 
\begin{align*}
R_{abcd}& =-\frac{1}{12}R(g_{ad}g_{bc}-g_{ac}g_{bd}) \\
R_{ab}& =\frac{1}{4}Rg_{ab}
\end{align*}%
where the Ricci scalar is $R=4\Lambda $. The field equations (\ref%
{fequations}) can now be reduced to the single equation 
\begin{equation}
\frac{1}{2}f-\Lambda =\Lambda f_{X}+2\Lambda ^{2}f_{Y}+\frac{4}{3}\Lambda
^{2}f_{Z}.
\end{equation}%
This equation must be satisfied by $f$ if the de Sitter universe is to be a
solution in any particular gravitational theory. This problem reduces to
that studied by Barrow and Ottewill \cite{BO} in the case $f=f(X)$. This
result establishes the situations where inflation of de Sitter sort can
arise from higher-order corrections to the gravitational lagrangian. In the
case where $f=f(X)$ alone it is appreciated that the resulting theory is
conformally equivalent to general relativity plus a scalar field \cite{cot,
maeda} with an asymmetric exponential potential and so either de Sitter or
power-law inflation is possible. Our results establish when de Sitter
inflation is possible in situations where the other invariants, $Y$ and $Z$
contribute to the Lagrangian, and the conformal equivalence with general
relativity is broken. The stability of the de Sitter solutions in these
theories will be studied elsewhere.

\section{Conclusions}

\label{conclusions}

We have analysed the existence conditions for solutions of G\"{o}del,
Einstein, and de Sitter type to exist in gravity theories derived from a
Lagrangian that is an arbitrary function of the curvature invariants $R$, $%
R_{ab}R^{ab}$ and $R_{abcd}R^{abcd}$. The existence conditions have been
systematically explored and a number of special choices of the Lagrangian
function that are of physical interest were worked out explicitly to display
the form of the solutions in terms of the Lagrangian properties. In the G%
\"{o}del case there is a simple condition on the metric parameters which
reveals whether or not time-travelling paths exist in the space-time. We
have evaluated this condition in general and explicitly in the examples
studied. A range of situations exist in which time-travel is either possible
(as in general relativity) or impossible depending on the form of the
gravitational Lagrangian function defining the theory. We find that it is
not possible to construct a theory, within this class, for which the
existence of closed time-like curves is forbidden for all perfect fluids.
The conditions for Einstein static and de Sitter solutions were also found.

\section{Acknowledgements}

TC is supported by the PPARC.

\section{Appendix: Derivation of equation (\ref{P})}

The variation of the action derived from integrating the density
(\ref{density}) over all space is
\begin{align}
\nonumber
\delta I_G &= \chi^{-1} \int d \Omega \sqrt{g} \Bigl[ \frac{1}{2} f g^{a b}
  \delta g_{a b} + f_X \delta X + f_Y \delta Y + f_Z \delta Z \Bigr]\\ 
\nonumber
&= \chi^{-1} \int d \Omega \sqrt{g} \Bigl[ \frac{1}{2} f g^{a b}
  \delta g_{a b} - f_X (R^{a b} \delta g_{a b} - g^{a b} \delta R_{a
  b}) \\
\label{var}
& \qquad \qquad \qquad \qquad \qquad - 2 f_Y (R^{c (a} {R^{b)}}_{c} \delta g_{a b} - R^{a b} \delta
  R_{a b}) - 2 f_Z ({R_{c d e}}^{(b} R^{a) e d c} \delta g_{a b} -
  {R_a}^{b c d} \delta {R^a}_{b c d}) \Bigr]
\end{align}
where use has been made of 
\begin{equation*}
\delta g^{a b} = -g^{a c} g^{b d} \delta g_{c d}
\end{equation*}
and
\begin{equation*}
\delta \sqrt{g} = \frac{1}{2} g^{a b} \delta g_{a b}.
\end{equation*}
The subscript $_G$ here denotes that we are
considering the gravitational part of the action only.  Using the relations \cite{DeW65}
\begin{equation*}
\delta R_{a b} = -\frac{1}{2} g^{c d} ( \delta g_{a b ; c d} + \delta
g_{c d ; a b} - \delta g_{a c ; b d} - \delta g_{b d ; a c})
\end{equation*}
and
\begin{equation*}
\delta {R^a}_{b c d} = \frac{1}{2} g^{a e} (\delta g_{e d ; b c} + \delta
g_{e b ; d c} - \delta g_{d b ; e c} - \delta g_{e c ; b d} - \delta
g_{e b ; c d} +\delta g_{c b ; e d})
\end{equation*}
we can then write
\begin{align*}
f_X g^{a b} \delta R_{a b} &\simeq -f_{X ; c d} (g^{a b} g^{c d} - g^{a
  c} g^{b d}) \delta g_{a b}\\
2 f_Y R^{a b} \delta R_{a b} &\simeq- \Bigl[ \square (f_Y R^{a b})+(f_Y
  R^{c d})_{; c d} g ^{a b} - 2 {{(f_Y R^{c (a})_;}^{b)}}_c \Bigr]
  \delta g_{a b}\\
2 f_Z {R_a}^{b c d} \delta {R^a}_{b c d} &\simeq 4 (f_Z R^{c (a b)
  d})_{; c d} \delta g_{a b}
\end{align*}
where $\simeq$ means equal up to terms which are pure divergences.  
Such terms are irrelevant here as they can be transformed via Gauss's
theorem to terms on the boundary which are assumed to vanish.
Substituting these expressions back into (\ref{var}) and making the
definition
\begin{equation*}
\delta I_G \equiv - \chi^{-1} \int d \Omega P^{a b}\delta g_{a b}
\end{equation*}
then gives equation (\ref{P}), which completes the derivation.


\begin{thebibliography}{99}
\bibitem{footnote1} As pointed out by De Witt \cite{DeW65} there is a fourth
possibility, namely $\epsilon^{a b c d} R_{e f a b} {R^{e f}}_{c d}$. However,
this contraction is of no physical interest due to parity considerations.
\end{thebibliography}
\end{document}